\documentclass[a4paper,11pt]{article}
\pdfoutput=1 

\usepackage{jheppub} 

\usepackage[T1]{fontenc} 

\usepackage{lmodern}
\usepackage{braket} 

\pdfoutput=1
\usepackage{epstopdf}

\usepackage{mathrsfs}
\usepackage{psfrag}
\usepackage{color}
\usepackage{hyperref}

\usepackage{slashed}
\usepackage{simplewick}
\usepackage{cancel}
\usepackage[utf8]{inputenc}

\usepackage[table]{xcolor}

\usepackage{amsmath,bbm,array,amsfonts,graphicx,wrapfig,arydshln,lscape,float,multirow,longtable,rotating,makecell}
\usepackage{url}

\usepackage{subcaption}

\usepackage[backgroundcolor=black!5, linecolor=green!60]{todonotes}

\hypersetup{pdftitle={},pdfcreator={},linkcolor=[rgb]{0.15,0.35,0.75},colorlinks=true,citecolor=[rgb]{0.675,0,0.2},urlcolor=[rgb]{0.15,0.35,0.65}}
\let\olditemize\itemize\renewcommand{\itemize}{\vspace{-2pt}\olditemize\setlength{\itemsep}{1pt}\setlength{\parskip}{0pt}\setlength{\parsep}{-0pt}}
\let\oldenumerate\enumerate\renewcommand{\enumerate}{\vspace{-4pt}\oldenumerate\setlength{\itemsep}{1pt}\setlength{\parskip}{0pt}\setlength{\parsep}{0pt}}

\newcommand*{\eps}{\varepsilon}

\newcommand*{\diff}{\mathop{}\!\mathrm{d}}

\newcommand*{\Diff}[1]{\mathop{}\!\mathrm{d^#1}}

\def\beq{\begin{equation}}
\def\eeq{\end{equation}}
\def\bsp#1\esp{\begin{split}#1\end{split}}
\newcommand{\be}{\begin{equation}}
\newcommand{\ee}{\end{equation}}
\newcommand{\bea}{\begin{eqnarray}}
\newcommand{\eea}{\end{eqnarray}}

\def\spa#1.#2{\left\langle#1\,#2\right\rangle}
\def\spb#1.#2{\left[#1\,#2\right]}
\newcommand{\e}{\epsilon}
\def\nn{\nonumber}

\def\cO{{\mathcal O}}

\renewcommand{\phi}{\varphi}

\def\eps{\epsilon}

\def\ksl{\not{\hbox{\kern-2.3pt $k$}}}

\def\e{\epsilon}
\def\eps{\epsilon}

\def\bom#1{{\mbox{\boldmath $#1$}}}

\def\spa#1.#2{\left\langle#1\,#2\right\rangle}
\def\spb#1.#2{\left[#1\,#2\right]}
\def\lor#1.#2{\left(#1\,#2\right)}
\def\sand#1.#2.#3{%
\left\langle\smash{#1}{\vphantom1}^{-}\right|{#2}%
\left|\smash{#3}{\vphantom1}^{-}\right\rangle}

\definecolor{airforceblue}{rgb}{0.36, 0.54, 0.66}
\definecolor{bananayellow}{rgb}{1.0, 0.88, 0.21}
\definecolor{bittersweet}{rgb}{1.0, 0.44, 0.37}
\definecolor{blue(ncs)}{rgb}{0.0, 0.53, 0.74}
\definecolor{bole}{rgb}{0.47, 0.27, 0.23}
\definecolor{brass}{rgb}{0.71, 0.65, 0.26}
\definecolor{bronze}{rgb}{0.8, 0.5, 0.2}
\definecolor{brgreen}{rgb}{0.0, 0.26, 0.15}
\definecolor{burgundy}{rgb}{0.5, 0.0, 0.13}
\definecolor{cherry}{rgb}{1.0, 0.72, 0.77}
\definecolor{cocao}{rgb}{0.82, 0.41, 0.12}
\definecolor{citrine}{rgb}{0.99, 0.82, 0.07}

\newcommand{\tred}[1]{\textbf{\textcolor{red}{#1}}}

\makeatletter\DeclareRobustCommand*{\bfseries}{\not@math@alphabet\bfseries\mathbf\fontseries\bfdefault\selectfont\boldmath}\makeatother

\title{\boldmath Deriving canonical differential equations for Feynman integrals
from a single uniform weight integral}

\author[\e]{Christoph Dlapa} 
\author[\e]{\hspace{-5pt}, Johannes Henn}
\author[\e]{and Kai Yan} 

\affiliation[\e]{Max-Planck-Institut f\"ur Physik, Werner-Heisenberg-Institut, D-80805 M\"unchen, Germany}

\emailAdd{dlapa@mpp.mpg.de}
\emailAdd{henn@mpp.mpg.de}
\emailAdd{kyan@mpp.mpg.de}

\preprint{MPP-2020-10}

\abstract{
Differential equations are a powerful tool for evaluating Feynman
integrals. Their solution is straightforward if a transformation to a canonical
form is found. In this paper, we present an algorithm for finding such a 
transformation. This novel technique is based on a method due to
H\"oschele et al.\ and relies only on the knowledge of a single integral
of uniform transcendental weight. As a corollary, the algorithm can
also be used to test the uniform transcendentality of a given integral.
We discuss the application to several cutting-edge examples, including
non-planar four-loop HQET and non-planar two-loop five-point integrals.
A Mathematica implementation of our algorithm is made available together with this paper.
}

\begin{document} 
\maketitle
\flushbottom

\newpage

\section{Introduction}

Feynman integrals are ubiquitous in perturbative quantum field theory. They are required in order to extract predictions from the theory beyond the leading perturbative order. As a consequence, they are important in many areas. A prominent example is collider physics, where the underlying scattering processes are described by Feynman diagrams, and consequently on-shell momentum space Feynman integrals are needed. Another example are off-shell position-space correlation functions (e.g. in conformal field theory), from which one can determine the scaling dimension of fields, or renormalization group coefficients.

Beyond their obvious importance for physics, Feynman integrals also have interesting connections to mathematics. 
The reason is that Feynman integrals are periods  \cite{Bogner:2007mn,Kontsevich2001},
and give rise to interesting classes of special functions.
Moreover, they can be studied using partial differential equations, and algebraic geometry plays an important role in the analysis of the integrands.

Given these motivations, it is not surprising that this has been an active area of research for decades, and continues to be.
It has produced numerous insights and new methods for Feynman integrals, as reviewed in 
\cite{Smirnov:2012gma,Henn:2014qga,Zhang:2016kfo,Kotikov:2018wxe}.
One important concept that has emerged, initially in the maximally supersymmetric theory $\mathcal{N}=4$ super Yang-Mills, is that functions of uniform transcendental (UT) weight play a special role. 
In the case of multiple polylogarithms, the weight refers to the (minimal) number of integrations that are needed to obtain the function, starting from a rational function. For example a logarithm has weight one, and a dilogarithm has weight two. 

The weight property serves as an important organizing principle, but also significantly simplifies calculations. It was observed that uniform weight integrals satisfy particularly simple, canonical differential equations. One of the present authors conjectured that one may always chose a basis of Feynman integrals such that their differential equations take a canonical form \cite{Henn:2013pwa}.
Indeed, it is known that any Feynman integral satisfies some $n$-th order (Picard-Fuchs) differential equation.
Equivalently, the latter can be rewritten as a coupled $n \times n$ first-order system of differential equations.
For example, let us assume one kinematic variable $x$ and take the dimension to be $\mathrm{D}=4-2 \epsilon$, then we have
\begin{align}\label{deonevariableintro}
\frac{\diff}{\diff x} \vec{f}(x,\epsilon) = A(x,\epsilon) \vec{f}(x,\epsilon)
\end{align}
In general for Feynman integrals (but under certain conditions on the basis vector $\vec{f}$), the matrix $A(x,\epsilon)$ is rational.
However, in general the dependence is very complicated, and hence solving such equations is a difficult task.
The statement of \cite{Henn:2013pwa} is that a basis transformation exists that  significantly simplifies the matrix to a 
form where the solution of the DE can be read off in terms of special functions.

Perhaps equally important to this guiding principle is the insight which Feynman
integrals evaluate to UT functions \cite{ArkaniHamed:2010gh}, and hence provide a suitable basis for the differential equations \cite{Henn:2013pwa}.
The key idea is that the loop {\it integrand} contains the necessary information, and that the latter can be extracted systematically by taking (multi-dimensional) residues.  
If no doubles poles are encountered along the way, and the maximal residues (that localize all integrations, and are called leading singularities) are constant, the corresponding Feynman integral is conjectured to be UT \cite{ArkaniHamed:2010gh}. 
See refs.\ \cite{Arkani-Hamed:2014via,Herrmann:2019upk} for further details.

Subsequently, a complementary approach \cite{Henn:2014qga,Lee:2014ioa} was proposed independently by one of the present authors, and by Lee. 
It consists in systematically constructing basis transformations that simplify $A(x,\epsilon)$ in eq.\ \eqref{deonevariableintro}. 
The first step focuses on simplifying the $x$-dependence, trying to make the expected singularity structure of the Feynman integrals manifest. 
The mathematics behind this has its roots in work by Moser \cite{Moser}, see also \cite{Wasow}.
In a second step, one simplifies the $\eps$ dependence. 
A complete algorithm was first given by Lee \cite{Lee:2014ioa}, and several computer programs implementing this and related ideas exist  \cite{Meyer:2016slj,Prausa:2017ltv,Gituliar:2017vzm}.

However, both methods have certain limitations. In the leading singularity approach, it is usually simple to find some UT integrals, but obtaining a complete set of basis integrals is more difficult. Also, sometimes it is necessary to analyze leading singularities beyond four dimensions \cite{Chicherin:2018old}. Similarly, in its current implementation the Moser approach works well only for  small matrices, and for few kinematic variables.

Here we present a different method that generalizes previous work by H\"oschele et\ al. \cite{Hoschele:2014qsa}.
The starting point is the knowledge that {\it one} integral is UT.
The key insight is that this assertion contains a lot of information (in some sense infinite amount
of information). We can use the latter to find, algorithmically, a complete UT basis, and hence obtain canonical differential equations. 

The reason this works can be understood as follows. The statement that an integral is UT in dimensional regularization means
that, at a given order $\epsilon^k$ in its Laurent expansion, the coefficient function has uniform transcendental weight $k$ (given some
choice for the overall normalization). In other words, we obtain a non-trivial constraint at each order in $\epsilon$. 
On the other hand, a single integral knows about the full set of integrals. Indeed, its derivatives generate (in general) all other function
in the integral basis. Our task is then to properly organize the information contained in the derivatives into UT functions as well. 
In this paper, we show how to do this algorithmically. 

As a result, starting from a single UT integral (and completing the basis in an arbitrary way), we derive algorithmically a transformation to a canonical form of differential equations (if such a transformation exists). 
As a corollary, this provides a test of the UT property of an integral.
This can be used to find the canonical form for differential equations for 
individual integral sectors, which is advantageous since this limits the size of then
matrices that are needed. At the same time, we wish to emphasize that our method
can also be used with just a single integral in the top sector as input, 
to transform the equations to canonical form in a single step. The applications we 
present in this paper suggest that this is feasible in practice even for large
systems of differential equations.

Our work considerably simplifies the application of the canonical differential equation method, as it is in general much easier to find
one UT integral, as compared to a full UT basis. Several methods and techniques exist for finding the integral to start from.
Moreover, conjecturally, loop corrections in $\mathcal{N}=4$ super Yang-Mills are given by uniform weight functions, and therefore this theory can be used as further inspiration for finding a good `seed' for generating the UT integrals needed
for any quantum field theory.

\section{Description of the method} 
\label{sec:method} 

\subsection{From the Picard-Fuchs equation of a uniform weight Feynman integral to a canonical system of first-order equations}

Feynman integrals are related to each other by integration-by-parts relations.
They can be reduced to a finite number of master integrals, which satisfy a system of homogeneous linear differential equations \cite{Kotikov:1990kg,Bern:1992em,Gehrmann:1999as,Argeri:2007up}. 
Alternatively, the coupled system of master integrals can be described by an $n$-th order differential equation, the so-called Picard-Fuchs equation, which defines the linear relation between a certain master integral and its higher-order derivatives.

In \cite{Henn:2013pwa}, it was proposed that a canonical basis of master integrals exists, 
which consists of integrals of uniform transcendental weight (UT). They satisfy a linear system of differential equations in a simple, canonical form: 
\begin{align} \label{decanonicalsinglevar}
\frac{\diff }{\diff x } \vec{g}(x,\epsilon) = \epsilon\,  B(x) \, \vec{g}(x,\epsilon) \,, 
\end{align}  
with $B(x)$ having only fuchsian singularities.
The solutions to the canonical equations \eqref{decanonicalsinglevar} can be constructed iteratively order by order in $\epsilon$.

Searching for a canonical basis of master integrals is an import topic in
higher-loop calculations of Feynman integrals.  
It becomes more challenging for large coupled systems where one needs to 
find all the linearly-independent UT integrals to bring the first-order  differential equations into canonical form.  
On the other hand, one can derive the Picard-Fuchs equation for any given master integral. 
The coefficients of the Picard-Fuchs equation have some special characteristics,
which allow to test if an integral is UT without referring to the other master integrals in the bases. 
More importantly, by establishing the relation between the Picard-Fuchs equation and the first-order differential equations,    
one can map between two bases that share one common integral. 
In this way we can construct a canonical basis $\vec{g}$ starting from an arbitrary basis $\vec{f}$ that contains one UT integral.  

Our starting point is one UT integral, which without loss of generality we denote by $f_1$.
We complete this to a basis of $n$ master integrals, denoted by $\vec{f}$.
Given the system of differential equations
\begin{align}  \label{desinglevargeneral}
\frac{\diff}{\diff x} \vec{f} =  A (x, \epsilon) \vec{f}   \,,
\end{align}
we can reformulate this in terms of the function $f_1$ and its derivatives only.
To this end, we take derivatives of eq. (\ref{desinglevargeneral}), 
\begin{equation}
\label{higherderivatives}
\begin{pmatrix}
 f'_1 & \cdots & f^{(n)}_1   \\   
 f_2' & \cdots   & f_2^{(n)}  \\
 \vdots & \ddots & \vdots \\ 
 f_n' &  \cdots   & f_n^{(n)} \\
 \end{pmatrix}    =  \left(  A^{[1]} \vec{f}, \ldots ,A^{[n]} \vec{f}  \right)  \,,
\end{equation}
with
\begin{align}
 A^{[1]} &:=  A \,, \\
   A^{[n]} &:= \frac{\diff}{\diff x} A^{[n-1]} + A^{[n-1]} A \quad\quad {\rm for} \; n>1 \,.
 \end{align}
As we are interested in the higher-order derivatives of $f_1$, we project 
this equation with the vector $\vec{v}_0 = (1,0,\ldots, 0)$ and define
\begin{align}
\label{defPsi}
\Psi :=  \left(  
 \begin{array}{c}  \vec{v}_0  A^{[1]}      \\   \vdots  \\  \vec{v}_0  A^{[n]}  \end{array} 
 \right)   .
\end{align}
Comparing to eq. (\ref{higherderivatives}), we see that the matrix $\Psi$ satisfies
\begin{align}  \label{2.3}
   (f_1', f_1'' , \ldots ,f_1^{(n)} )^{T}   =  \Psi (x, \epsilon)\,   \vec{f}  \,.
\end{align}   
Assuming that $  f'_1, \ldots  , f^{(n)}_1  $ are linearly independent\footnote{It may happen that only $m$ derivatives of $f_1$ are linearly independent, with $m<n$. We will discuss this special case separately.}, 
we can then invert the matrix $\Psi$, and write down an $n$-th order differential 
equation for $f_1$,   
\begin{align} 
& f_1  + \sum_{m=1}^n  b_m \, f_1^{(m)}  =0 \, ,   \label{eqPicardFuchs1} \\ 
& \text{where }    (b_1, \ldots ,b_n)   \equiv  - (1,0,\ldots, 0) \, \Psi^{-1}   \, .   \label{2.7} 
\end{align}
The coefficients $b_1, \ldots, b_n$ are rational functions  of $x, \epsilon$. 
By factoring out the denominators, the differential equation \eqref{eqPicardFuchs1}  takes a form where all  coefficients are polynomials (Pichard-Fuchs equation): 
\begin{align} 
\label{eqPicardFuchs2} 
\tilde{b}_0 \, f_1  + \sum_{m=1}^n  \, \tilde{b}_m f_1^{(m)}  =0 \, 
\end{align} 
Given that $f_1$ is a UT integral, let us point out that
 the $\epsilon$-degree of the coefficients of the Pichard-Fuchs equation 
must be bounded by a  certain finite number depending on the order of the equation.  
 The reason is the following:
Let  $p_m$ be the $\epsilon$-degree of the polynomial $\tilde{b}_m$,
   and  let us assume that 
 the maximum value of $ p_m + m  \; ( 0 \leq m \leq n )$ is $p_{\rm{max}}$.  
  Note that 
  the $m$-th derivative of $f_1$ is a sum of functions whose transcendental weights range between $-1$ and $-m$, i.e.\ $f_1^{(m)} = \sum_{k=1}^m  f_1^{(m,k)} $, where $m \geq 1$ and $ f_1^{(m,k)} $ has weight $-k$.  
After assigning  transcendental weight $-1$ to $\epsilon$,  
we group together terms in \eqref{eqPicardFuchs2}
that have equal  transcendental weight and thus obtain 
 $ p_{\rm  max} +1 $ independent  linear equations in the unknowns $\{ f_1, f_1^{(m,k)} \}$. 
  If non-trivial solutions exist, then $p_{\rm max} +1 $ must be less or equal to the total number of  unknowns, which is $ 1+ \sum_{m=1}^{n}  m $. 
 In particular, the equation of weight zero only involves $ f_1$, whose coefficient must vanish, and therefore $  \tilde{b}_0  = \cO (\epsilon )$.  
To conclude, we find the following necessary conditions for $f_1$ being a UT integral: 
 \begin{align}
 \label{UTcondition}
  p_m + m  \leq  \frac{ n (n+1)  }{2}, \quad    m =0, \ldots, n\,  \quad \text{and} \hskip.5cm \tilde{b}_0  = \cO (\epsilon )\, . 
 \end{align} 
These conditions  allow us to test whether  an integral is UT or not. Similar conditions were also discussed in \cite{Hoschele:2014qsa}.

Now let us propose a way of solving the differential system by the method of undetermined exponential function.  
First, we assume the existence of a canonical basis $\vec{g}$, 
where $g_1= f_1$ and all the other members of $\vec{g}$ are unknown.   
 $\vec{g}$ satisfies the canonical differential equations  \eqref{decanonicalsinglevar}  with unknown matrix $B(x)$, 
whose general solution is defined through an iterated integral 
\begin{align}
\label{equ:iteratedIntegralSol}
\vec{g} (x, \epsilon) =  
  \mathrm{P} \, e^{ \epsilon  \int_{x_0}^x B(x)\, \diff x }\,  \vec{g} (x_0, \epsilon) 
\end{align}
Here $ \int B(x) \,\diff x $ is the exponential function that needs to be determined. 
 As we shall see in the following section, 
 the linear relations among  $g_1 (= f_1)$ and its derivatives allows us to determine the matrix $B(x)$
 up to a constant similarity transformation. 
To start, 
we define a matrix $\Phi$ in complete analogy with the definition of $\Psi$ in eq. (\ref{defPsi}), but with $A$ replaced by $B$,
\begin{align}\label{defPhi}
\Phi :=  \left(  
 \begin{array}{c}  \vec{v}_0  B^{[1]}      \\   \vdots  \\  \vec{v}_0  B^{[n]}  \end{array} 
 \right)   .
\end{align}
The latter satisfies
\begin{align}  \label{2.9}
  (g_1', g_1'' , \ldots, g_1^{(n)} )^{T}  = \Phi (x, \epsilon)\,   \vec{g}  . 
\end{align}   
 Comparing eqs.  \eqref{2.3} and \eqref{2.9}, we see that the matrices $\Phi$ and $\Psi$  define the transformation matrix $T$ 
 between the two bases $\vec{f}$ and $\vec{g}$, 
\begin{align} 
\label{transformT}
 \vec{f} = T \, \vec{g} \, ,  \quad \text{where }  T \equiv  \Psi^{-1} \Phi \, . 
\end{align}
Knowing that $f_1 = g_1$,  the first line of $T$ or $T^{-1}$ must equal the unit vector $\vec{v}_0$.
Explicitly,
\begin{align}\label{constraintT}
\vec{v}_0  \Psi^{-1} \Phi   = \vec{v}_0.
\end{align}
Recall that $\Psi$ is explicitly known, while $\Phi$ depends on the unknown matrix $B$.
The constraint (\ref{constraintT}) will be important for finding $B$, and hence the transformation to the UT basis $T$.

\subsection{Ansatz for canonical differential equations, and determination of unknowns}
\label{sec:equsolve}

By assumption, the basis $\vec{g}$ is UT, and hence the differential equations are in canonical form (\ref{decanonicalsinglevar}).
We write the matrix $B$ appearing in that equation as
\begin{align} \label{eq-canonicalde-singlevar1}
B(x) = \sum_{l}  \frac{ \diff}{\diff x} a_l (x) \,   \mathbf{m}_l  \,,
\end{align}  
where the $\mathbf{m}_l$ are constant matrices. 
The $a_{l}$ are functions behaving as $ \diff a_l \sim \diff y/y$ near singular points $y(x)=0$. 
Their explicit form depends
on the Feynman integrals under consideration.
In the case of multiple polylogarithms\footnote{See section \ref{sec:conclusionandoutlook} for comments on generalization to elliptic polylogarithms and beyond.} 
they take the form
\begin{align} \label{eq-canonicalde-singlevar2} 
B (x) = \sum_{l}  \frac{ \diff}{\diff x} \log \alpha_l (x) \,   \mathbf{m}_l \,.
\end{align}  
This form makes it manifest that Feynman integrals have fuchsian singularities only.
The set of (rational or algebraic) functions $\alpha_l$ depends on the problem under 
consideration. 
For a given system of master integrals, one can read off the set of singularities from
the differential equations \eqref{desinglevargeneral}. In the case of a rational alphabet $\{ \alpha_l \}$
this is all we need. 
For instance, in most of the applications considered below, we have 
 \begin{align}\label{example3singularities}
B = \epsilon \left( \frac{1}{x} \mathbf{m}_{1} + \frac{1}{x+1} \mathbf{m}_2  + \frac{1}{x-1} \mathbf{m}_3 \right) \,.
  \end{align}
The precise form or number of singularities is not important for our method. 
%
%
For simplicity of notation and presentation, let us for the following assume the rational form \eqref{example3singularities}
with three singularities at finite distance.  
(These assumptions can be dropped  in special cases where we need different forms of ansatz for the alphabet, see discussions in section \ref{sec:specialcases}).

With this information, we see that eq.\ (\ref{constraintT}) effectively becomes
an equation for the constant ($\epsilon$- and $x$-independent) matrices $\mathbf{m}_l$, and products thereof\ (projected by the vector $v_{0}$).
We may profit from the fact that the matrices are constant by sampling the equations 
for different values of $x$. 
Moreover, the method can naturally be combined with finite field methods, see e.g. \cite{Klappert:2019emp,Peraro:2019svx}.

In summary, we have the constraint \eqref{constraintT}, together with the ansatz \eqref{eq-canonicalde-singlevar1} or \eqref{eq-canonicalde-singlevar2}
for the canonical differential equation matrix $B$. The latter is parametrized via the set of constant matrices $\mathbf{m}_{l}$, and determines $\Phi$, see eq.\ \eqref{defPhi}. 
More precisely, each row of $\Phi$  defines an $n$-dimensional vector, which is a polynomial in $\epsilon$, 
 \begin{align}
   \vec{\Phi}_m  (x, \epsilon) \equiv  \sum_{k =1}^m  \epsilon^k \, \vec{ \Phi}_m^k (x) \, , \quad  \vec{\Phi}_0  \equiv \vec{v}_0 .
 \end{align}
Substituting into \eqref{constraintT}, or equivalently \eqref{eqPicardFuchs2}, 
 we obtain a linear equation with $x$-dependent coefficients at each fixed order in $\epsilon$.
  At $\cO(\epsilon^k)$, the equation reads 
 \begin{align} 
 \label{eqPhik}
\sum_{r=1}^k  \sum_{m= r}^n   \tilde{b}_m^{k-r}  \hskip-.05cm (x)\,  \vec{ \Phi}_m^{r}  -  \tilde{b}_0^{k}  (x)\,  \vec{ \Phi}_0 =0\, ,
\quad  1\leq   k \leq p_{\rm max} \, ,
 \end{align}  
 where $\tilde{b}_m (x, \epsilon) \equiv  \sum_{k=0}^{p_m}  \epsilon^k\, \tilde{b}_m^{k} (x) $. 
 In order to solve for $\{ \vec{\Phi}_m^k \}$,  recall the definition \eqref{defPhi} with given ansatz \eqref{eq-canonicalde-singlevar2}, 
which implies 
 \begin{align} 
 \label{paraPhik}
 \vec{\Phi}_{m}^k &  =  \frac{\diff}{ \diff x }   \vec{\Phi}_{m-1}^k  
  +  \sum_{l}  \left[ \frac{\diff}{\diff  x}    \log \alpha_l (x) \right]    \vec{\Phi}_{m-1}^{k-1} \mathbf{m}_l     \,. 
 \end{align}
The recursive structure of  \eqref{paraPhik}
 allows us to solve \eqref{eqPhik} iteratively  and determine the constant matrices $\mathbf{m}_l$.  
Below we give details of this algorithm.

 Starting from $k=1$ and  assuming 3 letters in the ansatz, \eqref{eqPhik} and \eqref{paraPhik} lead to  
 \begin{align} 
 \label{eqPhi1} 
   \sum_{l=1}^3  \sum_{m= 1}^n   \tilde{b}_m^{0}  (x)
    \left[ \frac{\diff^m}{\diff  x^m}    \log \alpha_l (x) \right]   \,  \vec{v}_0 \mathbf{m}_l    -  \tilde{b}_0^{1}  (x)\,  \vec{v}_0 =0.
 \end{align}  
 Assuming the $\alpha_l(x)$ are rational, \eqref{eqPhi1} can be sampled for a set of different generic numbers which gives rise to 
 a system of algebraic equations   linear in  $\vec{v}_0 \mathbf{m}_l $.  
 %
 Solving for the three unknowns, we find
\begin{align} 
\label{slvPhi1}
\vec{v}_0 \mathbf{m}_l  = \sum_{j=0}^{3-r_1}  \beta_{j}^l\,  \vec{v}_j  \, , \quad   l =1,2,3
\end{align}
where $r_1$ is the rank of the linear system \eqref{eqPhi1} sampled over a set of numbers and  $ \beta_{j}^l $ are rational numbers.  
Let $N_0 = 1, N_1 = 3 N_0  - r_1$. The solutions  in  \eqref{slvPhi1}  introduce $N_1$ independent free vectors $\vec{v}_1, \ldots, \vec{v}_{N_1}$ (e.g. one could designate $N_1$ unknowns as free vectors that parametrize all the other unknowns). \eqref{slvPhi1}  defines a set of linear relations 
  which, applied to \eqref{paraPhik}, reduce $\vec{\Phi}_m^{1}$ and
 $\vec{\Phi}_m^{2}$  to linear combinations of $ \{  \vec{v}_0, \vec{v}_i , \vec{v}_i \mathbf{m}_l \} ,  \, i = 1, \ldots, N_1$. 
 Then we can repeat the analysis for \eqref{eqPhik}  at $k=2$,  
 treating $\vec{v}_i \mathbf{m}_l$ as the unknowns.  

To generalize, the algorithm introduces a total number of $S_{k-1} \equiv \sum_{j=1}^{k-1} N_j$ free vectors through
 step $1\, {\rm to} \,k-1$, $(k \geq 2)$.   
 Moving on to step $k$,  we first substitute the solutions obtained at the previous step into \eqref{paraPhik} and reduce  $ \vec{\Phi}_m^k$ onto a set of vectors $Q_k \equiv \{ \vec{v}_0 , \; \vec{v}_1, \ldots, \vec{v}_{S_{k-1}} , \;  \vec{v}_{S_{k-2}+1} \mathbf{m}_{l} , \ldots,  \vec{v}_{S_{k-1}} \mathbf{ m}_l  \}$, 
  which includes $S_{k-1} $ free vectors
$\vec{v}_i$ and $3 N_{k-1}$ unknown vectors $ \vec{v}_{i} \mathbf{m}_l$. 
Evaluating \eqref{eqPhik} over a set of generic numbers then leads to a linear system of equations in $Q_k$ of rank $r_k$. 
 Solving for the unknowns, we obtain  $3 N_{k-1}$ independent  relations,
\begin{align} 
\label{slvPhik}
\hskip+.5cm \vec{v}_i \mathbf{m}_l  = \sum_{j=0}^{ S_k }   \beta_{ij}^l\,  \vec{v}_j ,  \quad \hskip+.2cm  \text{where}  \;   S_{k-2} <   i  \leq  S_{k-1} ,  \quad l =1,2,3 \,
\end{align} 
which introduce $N_k$ new free vectors. $ N_k \equiv S_k - S_{k-1} = 3 N_{k-1} - r_k $. Again, $\beta_{ij}^l$ are rational constants. 
Since the algorithm constructs linear equations step-by-step, 
 the relations in \eqref{slvPhik} must be 
 independent from those obtained at previous steps. 
 Iterating the above procedure until we reach a certain step where $N_k =0$, and $S_k = S_{k-1} \equiv |S|-1$, the algorithm terminates. 
 Each row of the $\Phi-$matrix is now completely determined as a linear combination of vectors in 
 $Q \equiv \{ \vec{v}_0 , \vec{v}_1 ,\ldots , \vec{v}_{|S|-1} \}$. 
Solutions in all previous steps  
combine into $ 3 |S|$ independent linear relations $ Q\, \mathbf{m}_l = \beta^l Q  \,, $ 
where $\beta^l \, (l =1, 2,3) $ are three $|S| \times |S|$ matrices, whose components  $\beta^{l}_{ij}$ are rational numbers.

In fact, if non-trivial solutions exist, then we must have  $|S| =n$,  the rank of the coupled linear system.
 If $|S|$ is greater than $n$, some of the vectors  in $Q$ are not linearly independent,  which contradicts the claim that the $3 |S|$ linear relations we obtained are independent. 
If $|S|$ is smaller than $n$, then the rank of $\Phi$-matrix is smaller than $n$, which contradicts our assumption that $f_1', \ldots, f_1^{(n)}$ are linearly independent.  Therefore $|S|=n$, $Q$ is an $n \times n$ invertible constant matrix and $ \mathbf{m}_l = Q^{-1} \beta^{l} Q$.  

Finally, we need to make some explicit choice for the vectors in $Q$. For example, we could choose $\vec{v}_i$ to 
be the $(i+1)$-th row of the identity matrix, such that the $\mathbf{m}_l=\beta^l$.  
 A different choice of $\vec{v}_i$ generates a constant linear transformation $P$ acting on $Q$, which preserves its first component. 
Correspondingly, the matrices $\mathbf{m}_l$ differ by  a constant similarity transformation  
\begin{align}
\label{constantLT}
Q \mapsto Q P:\qquad \vec{v}_0 P=\vec{v}_0, \qquad \mathbf{m}_l = Q^{-1} \beta^{l} Q  \mapsto P^{-1} \mathbf{m}_l P  .
\end{align} 

To summarize,  the constraint \eqref{constraintT}  suggests a system of linear equations \eqref{eqPhik}.
 We develop an algorithm to solve them iteratively  order-by-order in $\epsilon$.  
Given the assumption that $f_1$ is UT,  the $\Psi$-matrix has rank $n$ and the ansatz for  letters $\alpha_l$ is complete,  the algorithm will terminate at a certain order in $\epsilon$, when it finds
 a non-trivial solution for the $\mathbf{m}_l$  that determines $\Phi(x, \epsilon)$ and hence $T(x, \epsilon)$,  up to a constant linear transformation.   
 Equations at higher orders in $\epsilon$ must be trivially satisfied. 
 We provide a Mathematica implementation of this algorithm, see section \ref{sec:MathImplement}.

%
%
%
%
%
%

\subsection{Generalization to multi-variable case}

Unlike methods based on the Moser algorithm, see e.g.\ \cite{Henn:2014qga,Lee:2014ioa,Meyer:2016slj,Prausa:2017ltv,Gituliar:2017vzm}, 
the inclusion of multiple scales does not pose a significant problem within our approach. 
The main point is that the canonical form of the differential equations for multiple polylogarithms
still only depends on a finite number of constant matrices $\mathbf{m}_{l}$ that are determined by our 
procedure. Concretely, the multi-variable generalization of eq.\ (\ref{eq-canonicalde-singlevar2}) is
\begin{align}\label{conclusion-multi-variable}
\diff {\vec g}({ \vec{x}},\epsilon) = \epsilon \left[ \sum_{l} \mathbf{m}_{l}\, \mathrm{dlog}(\alpha_{l}({ \vec{x}})) \right]  {\vec g}({ \vec{x}},\epsilon) \,.
\end{align}
Here $\vec{x} = \{ x_1, \ldots, x_m \}$ denotes a set of variables, $\diff = \sum_{i} \diff x_{i} \partial_{x_i}$, and the set $\{ \alpha_{l}({x}) \}$ is the alphabet. 
Although the whole differential depends on multiple variables $\vec{x}$,
one may always view it as a single-variable problem by 
treating all but one preferred variable (e.g. $x_1$) as constants. 
The knowledge about  the $x_1$-dependent letters provides sufficient information to reconstruct the answer.  
Let us  illustrate here how the method works in the two-variable case. 
We will present  a state-of-the-art multi-variable example in section \ref{sec:fivevariable}. 

Starting with a system of differential equations for a basis $\vec{f}$ containing one UT integral
\begin{align}
\label{detwovariable}
\frac{\partial}{\partial x } \vec{f} = A_x(x, y, \epsilon ) \vec{f}, \quad   \frac{\partial}{\partial y }  \vec{f} = A_y(x, y, \epsilon ) \vec{f}\,, 
\end{align}
we make an ansatz for the alphabet $\{ \alpha_l(x,y) \}$, 
 and hence  the differential equation for a UT basis $\vec{g}$ in canonical form: 
\begin{align}
\label{ansatz-twovariable}
\diff \vec{g} = \diff B (x, y, \epsilon) \, \vec{g} = 
 \epsilon \sum_{l}   \mathbf{m}_l\, \mathrm{dlog} \alpha_l(x,y) \,
\end{align} 
We now study the Pichard-Fuchs equation in $x$, treating $y$ as a constant $y=y_0$.
For convenience, we assume that the first $L$ letters depend on $x$ (and possibly $y$) and the others depend only on $y$.
Using the short-hand notation $ B_x \equiv \frac{\partial}{ \partial x} B $, the partial derivative $B_x$ only depends on $\mathbf{m}_1, \ldots, \mathbf{m}_L$. 
Similar to the one-variable case (see \eqref{defPsi}), 
by taking partial derivatives we define the matrix $\Psi_x$ through $A_x$
 and likewise $\Phi_x$ through $B_x$. 
They satisfy the following relations, 
\begin{align}  
\label{defPhixy}
  \Big(  \frac{\partial}{ \partial x } f_1,  \frac{\partial^2}{ \partial x^2} f_1 , \ldots ,\frac{\partial^n}{ \partial x^n}   f_1  \Big)^{T}  & = \Psi_x(x,y, \epsilon) \vec{f}  \nn \\
  & =  \Phi_x (x, y, \epsilon)\,   \vec{g} \,,
\end{align}   
which implies the constraint
\begin{align} 
\label{constraintTpax}
 \vec{v}_0  \, \Psi^{-1}_x \, \Phi_x = \vec{v}_0 \,. 
\end{align}  
The next step is to solve this system of constraints for $\mathbf{m}_{l} \,  ( l = 1, \ldots, L)$. as described in the previous section, solving \eqref{constraintTpax}  order-by-order in $\eps$. 
We will obtain a set of linear relations \eqref{msolution},  where  $\beta^l \;  (l = 1, \ldots, L) $  are $n \times n$ rational matrices 
 and $Q$ denotes $\{ \vec{v}_0 , \ldots, \vec{v}_{n-1} \}$, 
 \begin{align}
 \label{msolution}
 Q \mathbf{m}_{ l} =  \beta^l Q 
 \end{align} 
For any given $Q$,  this defines our solution for $\mathbf{m}_{l}$. 
 Let us emphasize that  the solution is constant and, in particular, independent of $y_0$. 
This is because, by construction, 
the coefficients $\beta^l$ obtained by the algorithm are unique (the algorithm runs with a specific set of $\{ \vec{v}_1 , \ldots, \vec{v}_{n-1} \}$ chosen from $\{ \vec{v}_0\mathbf{m}_{i},   \vec{v}_0\mathbf{m}_{i}\mathbf{m}_{j}, \ldots  \}$). 
If there exists a $y$-independent solution for $\mathbf{m}^l$, then it must satisfy \eqref{msolution} at $y=y_0$, and we have found it. 
These constant matrices $\mathbf{m}_l$, together with
the knowledge about the first $L$ letters in $\{ \alpha_l \}$,
allow us to determine the $x$- and $y$-dependent matrix $B_x$ and hence $\Phi_x$.  
In this way we find the transformation matrix $T(x,y, \epsilon) \equiv \Phi_x^{-1} \Psi_x $, which brings the partial 
differential matrix $A_x$  into canonical form. 

Now we will argue that $\vec{g} = T \vec{f} $ is a UT basis.  Let us assume there exists a UT basis $\vec{g}_{\rm UT} $, whose first component is $f_1$ and which satisfies the differential equation  $ \frac{\partial}{d x}   \vec{g}_{\rm UT} =   \widetilde{B}_x \,  \vec{g}_{\rm UT}$. 
From the analysis in the previous section (see \eqref{constantLT}),  
solutions the $\mathbf{m}_l$ are related by a similarity transformation which leaves its first row invariant.  
Therefore, there exists a constant transformation matrix $P$ between $\widetilde{B}_x$ and our solution for $B_x$, such that  
 \begin{align}
 \widetilde{B}_x = P^{-1} B_x P , \quad  \vec{v}_0 P = \vec{v}_0 \,. 
 \end{align}
 Following from definition \eqref{defPhi}, the matrix $\widetilde{\Phi}_x$ defined through $\widetilde{B}_x$ is related to $\Phi_x$ through a constant linear transformation 
 \begin{align}
 \widetilde{\Phi}_x = \Phi_x P  \,. 
 \end{align}
 The same constant matrix transforms the UT basis $\vec{g}_{\rm{UT}}$ into $\vec{g}$  
 \begin{align}
  \vec{g}& =  \Phi_x^{-1}    
  \Big(  \frac{\partial}{ \partial x } f_1,  \frac{\partial^2}{ \partial x^2} f_1 , \ldots, \frac{\partial^n}{ \partial x^n}   f_1  \Big)^{T}    \nn \\
 & =   P\, \widetilde{\Phi}_x^{-1} \Big(  \frac{\partial}{ \partial x } f_1,  \frac{\partial^2}{ \partial x^2} f_1 , \ldots, \frac{\partial^n}{ \partial x^n}   f_1  \Big)^{T} =  P \, \vec{g}_{\rm UT} .  
 \end{align}
Therefore $\vec{g}$ itself  is a UT basis.   
 The above argument holds as long as the $x$-derivatives for $f_1$ couple to all master integrals in the family, 
 such that $\Phi_x$ and $\Psi_x$ are invertible. 
 In the process of searching for UT integrals,  the algorithm refers only to the partial derivatives in $x$, thus becoming very efficient. The algorithm does not require a prior knowledge of the complete set of letters. 
 In practice,  the $x$-independent letters can be determined afterwards  by  transforming  \eqref{detwovariable}  into the whole differential in canonical form.

\subsection{Special cases with degenerate \texorpdfstring{$\Psi$}{Psi}-matrix and algebraic letters}
\label{sec:specialcases}

In previous sections we explained how the algorithm applies to standard one- and two-variable examples.
For simplicity of presentation, we assumed
\begin{enumerate}
\item  One can find a UT integral $f_1$ whose higher-order derivatives couple to all master integrals in the system. \label{pt1}
\item   The differential equations contain only rational letters, and hence takes the form of  \eqref{example3singularities}. \label{pt2}
\end{enumerate}
In reality,  these assumptions can be dropped when needed.
 Now we will discuss the subtleties of applying our algorithm when  \ref{pt1}) and \ref{pt2}) no longer hold.

Regarding assumption \ref{pt1}),
 we would like to comment on the situation where
 derivatives of the first UT integral only couple to a subset of master integrals. 
For an $n \times n$ coupled system, depending on our choice of $f_1$, 
the number of its  independent higher-order derivatives could be less then $n$. 
The corresponding $\Psi$-matrix is then degenerate. 
Typically this could happen when $f_1$ belongs to a  sub-topology of the integral family. 
Sometimes it occurs even if $f_1$ is in the top sector.
 One example would be the scalar integral in three-loop ladder integral family, 
 whose derivatives only couple to 23 out of a total number of 26 master integrals, see section \ref{sec:fourloop}.  
In these situations the algorithm still works, but we need to search for a second UT integral $f_2$, such that the union of independent derivatives of $f_1$ and $f_2$ couples to all master integrals.
One way to proceed can then be to first bring a linearly independent sub-block of the differential equation into canonical form, and then use the second UT integral to work on the remaining part.

Another very efficient approach can be to use both UT integrals simultaneously:
For example, one can find a set of linearly independent derivatives $\{f_1' , \ldots, f_1^{(n_1)} , f_2' , \ldots, f_2^{(n-n_1)} \}$, which contains $n_1$ and $n- n_1$ higher order derivatives of $f_1$ and $f_2$, respectively. 
They form a basis of the coupled system.   
Starting from a master integral basis $\vec{f} \equiv (f_1 , f_2, \ldots, f_n)^T$,  we then construct a matrix $\Psi$ in the same way as in \eqref{defPsi}:
\begin{align}
\label{twoUTPsi}
\Psi :=  \left(  
 \begin{array}{c}  \vec{v}_0  A^{[1]} \\   \vdots       \\ \vec{v}_0 A^{[n_1]} \\   \vec{v}_1  A^{[1]} \\ \vdots \\ \vec{v}_1 A^{[n-n_1]}  \end{array} 
 \right)   
\end{align}
where $\vec{v}_0 , \vec{v}_1$ are the first and second row of the identity matrix.
The $\Psi$-matrix thus defined is invertible and satisfies the relation
\begin{align}  
\label{twoUTfdev}
   (f_1',  \ldots ,f_1^{(n_1)} , f_2', \ldots, f_2^{(n-n_1)} )^{T}   =  \Psi (x, \epsilon)\,   \vec{f}  \,.
\end{align}   
Given the two UT integrals that are already known, we can search for a UT basis $\vec{g} = ( g_1, g_2 , \ldots, g_n )^T$, where $g_1 = f_1, g_2 = f_2$. 
As we discussed before, $\vec{g}$ satisfies \eqref{decanonicalsinglevar}, and therefore
\begin{align} 
\label{twoUTgdev}
 (g_1',  \ldots ,g_1^{(n_1)} , g_2', \ldots, g_2^{(n-n_1)} )^{T}   =  \Phi (x, \epsilon)\,   \vec{g}  , 
\end{align}
where $\Phi$ is defined in the same way as $\Psi$ in $\eqref{twoUTPsi}$ with $A$ replaced by $B$. 
\eqref{twoUTfdev} and \eqref{twoUTgdev} imply  the following system of constraints, 
\begin{align}
\label{twoUTconstraintT}
\left(  
 \begin{array}{c}  \vec{v}_0      \\   \vec{v}_1   \end{array} 
 \right)    \Psi^{-1} \Phi =
  \left(  
 \begin{array}{c}  \vec{v}_0      \\   \vec{v}_1   \end{array}
 \right)  \;,
\end{align}
By solving these constraints, we find the transformation between the two bases $\vec{f}$ and $\vec{g}$.  

Next, we will come to assumption \ref{pt2}) about rational letters. 
 In the cases where \ref{pt2}) holds, 
 we should allow the ansatz for the $B$-matrix to take a more general form compared with the oversimplified version in \eqref{example3singularities}. 
In particular, the differential equations might contain fuchsian singularities at zeros of  a certain higher-degree polynomial,
 which can be algebraic and complex numbers (e.g.  sixth-root of unity).  More generally, we can assume a factorized form with $k$ singularities, 
\begin{align}
B = \epsilon  \sum_{l=1}^{k}  \frac{\mathbf{m}_l}{x-x_{l}} \,,
\end{align}
where the $x_i$ are roots of a  degree-$k$ polynomial $P_k (x)$ with real and rational coefficients.
For the purpose of  analyzing  equations  \eqref{eqPhik} by the finite-field method, 
 we need to avoid writing down an ansatz for the $\Phi$-matrix
that explicitly  contains algebraic numbers.
Thus,  it is advantageous to make the ansatz for $B$ in the following form, 
\begin{align}
\label{examplerootsingularities}
B = \epsilon  \sum_{l=1}^{k}  \frac{ x^{l-1} }{P_k (x)} \mathbf{m}_l \,.
\end{align}
Now that  \eqref{examplerootsingularities} guarantees that only rational functions and numbers appear in \eqref{eqPhik},  we expect to find solutions for the constant matrices  $\mathbf{m}_l$ whose elements are real and rational numbers.   
We can again use  finite fields to search for such solutions, as explained in section \ref{sec:equsolve}.

So far we have limited our discussion to differential equations with rational letters, 
where one can always find a rational transformation matrix $T(x,\epsilon)$ that brings  \eqref{desinglevargeneral}  to canonical form. 
There are also situations where a rational transformation does not exist and hence square roots are required.  
For rational alphabets, the individual alphabet letters can be read off from the original differential equation.
The case of algebraic dependence on the kinematics is more subtle, as the latter are not immediately
apparent. However, studying the singular behavior of the differential equation (in particular, the critical
exponents in each limit) provides this additional information. 
For example, if the differential equation \eqref{desinglevargeneral} is in fuchsian form, we can expand it around any
singular point $x= x_i$, 
\begin{align}
A(x, \epsilon )  =  \frac{ B_i (\epsilon) } {x - x_i } + \mathcal{O} \big(  (x- x_i)^0 \big) 
\end{align} 
The exponential of the coefficient matrix $B_i$ determines the 
asymptotic behaviour of the solution to \eqref{desinglevargeneral}  as $x \rightarrow x_i$, 
\begin{align}   
 \vec{f} (x, \epsilon)  \sim  (x- x_i)^{B_i (\epsilon) } \vec{f}_0 (\epsilon) \,. 
\end{align} 
At $\epsilon=0$, 
 if  $B_i$ contains non-integer eigenvalues, they cannot be transformed away 
via the so-called balance transformation \cite{Lee:2014ioa}. 
 Put differently,  the square roots  in $(x- x_i)^{B_i}$ cannot be removed by a rational linear  transformation on the basis,   
and therefore $\sqrt{x- x_i}$ must  be included in our ansatz for the alphabet. 
Repeating the analysis for all fuchsian singular points should allow us to  
find all the square-root letters in the alphabet.  In many cases,  it is possible to find a set of variables that rationalizes all letters, (see e.g.\ \cite{Besier:2019kco}).  After such a change of variables the problems simplify into those with only rational letters. 
In this way we apply the algorithm to work out a four-variable example with square-root letters, as we will demonstrate in \ref{sec:fivevariable}.

 
Let us emphasize that the method of our algorithm should work  for algebraic letters without  rationalization. 
In this case the Pichard-Fuchs equation will still be rational, but the $\Phi$-matrix, and hence the coefficients of  system of linear equations for the $\mathbf{m}_l$, will be algebraic. 
Nevertheless, the constant $\mathbf{m}_l$ matrices  will not depend on the square roots. 
One may still apply (a variation of) our algorithm to search  for rational solutions
and provide a quick test of the conjectured  algebraic alphabet letters. 

\section{Examples and applications}
\label{sec:applications}

In this section we present examples and applications of our algorithm.

We use relatively standard notation for Feynman integrals in the context of differential equations.
The integral families are defined as
\begin{equation}\label{deffamily}
 G_{a_1,\ldots,a_n}=e^{L\epsilon\gamma_E}\int\frac{\Diff{D}k_1\ldots\Diff{D}k_L}{(i\pi^{\mathrm{D}/2})^L}\frac{1}{D_1^{a_1}\ldots D_n^{a_n}},
\end{equation}
where $L$ is the number of loops, $\gamma_E$ is the Euler-Mascheroni constant, the denominator factors $D_i$ are defined in the respective subsections and we are using the mostly minus metric $(+-\cdots-)$.

In subsection \ref{sec:threeloop} we apply the algorithm to two three-loop four-point integral families that first have been computed in ref.\ \cite{Henn:2013fah}. In subsection \ref{sec:fourloop} we present a new result for a four-loop four-point integral and in subsection \ref{sec:cusp} we bring a non-planar four-loop sector that appears in the computation of the angle-dependent cusp-anomalous dimension into canonical form. Finally, as a multi-variable example, we apply the algorithm to the top sector of a non-planar two-loop five-point family which was computed in \cite{Chicherin:2018old,Abreu:2018aqd}. A summary of the performance of our implementation on a standard desktop computer with twelve logical cores can be found in table \ref{tab:benchmarks}.

The IBP reductions necessary for computing the initial differential equations were done either by the integral reduction program \textsc{FIRE6} \cite{Smirnov:2019qkx} or by using the tools available in the \textsc{FiniteFlow} framework \cite{Peraro:2019svx}. Both codes depend on \textsc{LiteRed} \cite{Lee:2013mka} to generate the IBP identities.

\subsection{Full differential equation for planar three-loop integrals}
\label{sec:threeloop}

\begin{figure}[t]
\centering
\begin{subfigure}[t]{0.30\textwidth}
	\begin{center}
	\includegraphics[width=1\columnwidth]{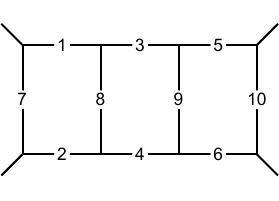}
		\caption{}
		\label{fig:P3L4P-a}
	\end{center}
\end{subfigure}
\hspace{2cm}
\begin{subfigure}[t]{0.30\textwidth}
	\begin{center}
	\includegraphics[width=1\columnwidth]{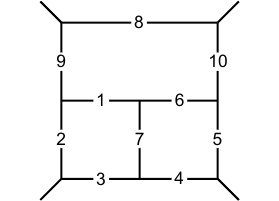}
		\caption{}
		\label{fig:P3L4P-b}
	\end{center}
\end{subfigure}
\caption{Planar three-loop four-point integrals. The number of MI is 26 for (a) and 41 for (b).}
\label{fig:P3L4P}
\end{figure}

As a first example, we apply the algorithm to the two three-loop four-point integral families shown in figure \ref{fig:P3L4P}.
The definition of the factors in eq.\ \eqref{deffamily} is
\begin{equation}
 \begin{aligned}
 D_1&=-k_1^2, &D_2&=-(k_1+p_1+p_2)^2, &D_3&=-k_2^2,\\
 D_4&=-(k_2+p_1+p_2)^2,\quad &D_5&=-k_3^2,\quad &D_6&=-(k_3+p_1+p_2)^2,\\
 D_7&=-(k_1+p_1)^2,\quad &D_8&=-(k_1-k_2)^2,\quad &D_9&=-(k_2-k_3)^2,\\
 D_{10}&=-(k_3-p_3)^2,\quad &D_{11}&=-(k_1-p_3)^2,\quad &D_{12}&=-(k_2+p_1)^2,\\
 D_{13}&=-(k_2-p_3)^2,\quad &D_{14}&=-(k_3+p_1)^2,\quad &D_{15}&=-(k_1-k_3)^2
 \end{aligned}
\end{equation}
and
\begin{equation}
 \begin{aligned}
 D_1&=-(k_1-k_3)^2, &D_2&=-(k_1+p_1)^2, &D_3&=-(k_1+p_1+p_2)^2,\\
 D_4&=-(k_2+p_1+p_2)^2,\quad &D_5&=-(k_2-p_3)^2,\quad &D_6&=-(k_2-k_3)^2,\\
 D_7&=-(k_1-k_2)^2,\quad &D_8&=-k_3^2,\quad &D_9&=-(k_3+p_1)^2,\\
 D_{10}&=-(k_3-p_3)^2,\quad &D_{11}&=-(k_3+p_1+p_2)^2,\quad &D_{12}&=-(k_2+p_1)^2,\\
 D_{13}&=-(k_1-p_3)^2,\quad &D_{14}&=-k_1^2,\quad &D_{15}&=-k_2^2
 \end{aligned}
\end{equation}
for integral families \ref{fig:P3L4P}(a), and \ref{fig:P3L4P}(b), respectively.
The top sector is defined by $G_{1,1,1,1,1,1,1,1,1,1,0,0,0,0,0}$ in both cases.

The integrals were computed previously in ref.\ \cite{Henn:2013fah} with the differential equations method.
In this case it is relatively straightforward to find a complete UT basis as in \cite{Henn:2013fah}, 
or even a complete dlog integrand basis \cite{WasserMSc}. Nevertheless, we find it instructive to benchmark our new method using these sets of integrals. We will see that, for each integral family, a single UT integral from the top sector is sufficient to derive the full  canonical differential equation. The corresponding matrices are of size $26 \times 26$ and $41 \times 41$, respectively.
A suitable initial integral is easily found using \cite{WasserMSc}, or by taking inspiration from the 
perturbative expansion of $\mathcal{N}=4$ super Yang-Mills \cite{Bern:2005iz}.

Concretely, we took the following integrals as our starting point,
\begin{align}
f_1=g_1=\eps^6 G_{1,1,1,1,1,1,1,1,1,1,-1,0,0,0,0}
\end{align}
and
\begin{align}
f_1=g_1=\eps^6x^2G_{1,1,1,1,1,1,1,1,1,1,-1,0,0,0,0}
\end{align}
for integral families \ref{fig:P3L4P}(a), and \ref{fig:P3L4P}(b), respectively. We completed them to basis by taking linearly independent integrals suggested by the integral reduction programs. In other words, no optimization was done on the other integrals.

The integrals depend on one dimensionless variable $x=t/s$.  See \cite{Henn:2013fah}  for more details.
The differential equation matrix $A(x)$ has the singular points $x=0, -1, \infty$, and consequently we take
the alphabet in eq.\ (\ref{eq-canonicalde-singlevar2}) to be $\vec{\alpha} = \{ x, 1+x \}$.
With this as input, our algorithm effortlessly found the transformation matrix $T$, see table \ref{tab:benchmarks}.

\subsection{New result for a four-loop four-point integral}
\label{sec:fourloop}

\begin{figure}[t]
\centering
	\begin{center}
	\includegraphics[width=0.3\columnwidth]{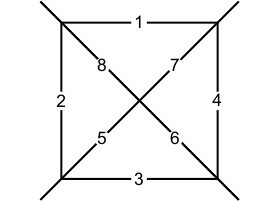}
    \end{center}
\caption{Planar four-loop four-point integral. The number of MI is 19.}
\label{fig:P4L4P}
\end{figure}

Let us now present an application to previously unknown four-loop integrals.
The definition of the integral family is
\begin{equation}
 \begin{aligned}
 D_1&=-k_4^2, &D_2&=-(k_1+p_1)^2, &D_3&=-(k_2+p_1+p_2)^2,\\
 D_4&=-(k_3+p_1+p_2+p_3)^2, &D_5&=-(k_1-k_2)^2, &D_6&=-(k_2-k_3)^2,\\
 D_7&=-(k_3-k_4)^2, &D_8&=-(k_1-k_4)^2, &D_9&=-k_1^2,\\
 D_{10}&=-(k_2+p_1)^2, &D_{11}&=-(k_3+p_1+p_2)^2, &D_{12}&=-(k_4+p_1+p_2+p_3)^2,\\
 D_{13}&=-k_2^2, &D_{14}&=-k_3^2, &D_{15}&=-(k_1-k_3)^2,\\
 D_{16}&=-(k_1+p_1+p_2)^2, &D_{17}&=-(k_1+p_1+p_2+p_3)^2, &D_{18}&=-(k_2-k_4)^2,\\
 D_{19}&=-(k_2+p_1+p_2+p_3)^2, &D_{20}&=-(k_3+p_1)^2, &D_{21}&=-(k_4+p_1)^2,\\
 D_{22}&=-(k_4+p_1+p_2)^2. & & & &
 \end{aligned}
\end{equation}
The sector shown in figure \ref{fig:P4L4P} is $G_{1,1,1,1,1,1,1,1,0,0,0,0,0,0,0,0,0,0,0,0,0,0}$ and, together with the subsectors, there are 19 MI. 

This case is particularly interesting for the following reason. 
After taking certain residues, the scalar integrand exhibits a double pole. This can
be understood from a power counting argument and comes from the fact that the integral has relatively few propagators.
As a consequence, the scalar integral, or integrals with the same propagator structure, and numerators, do
not have a {\it dlog} form in four dimensions. There are possible remedies to this, including writing down
a {\it dlog} basis for the  `top'  sector shown in figure \ref{fig:P4L4P}, and then solving for the full family
of integrals. 

Instead, here we wish to use the new tool for testing UT integrals to find directly UT integrals in the sector of figure \ref{fig:P4L4P}.
We proceed as follows. First, we look directly at the scalar integral $G_{1,1,1,1,1,1,1,1,0,0,0,0,0,0,0,0,0,0,0,0,0,0}$. Although this turns out to be a suitable integral (it is UT up to an overall normalization in $\eps$), only 18 of its derivatives are linearly independent. One way to proceed can be to first bring an $18\times18$ block of the matrix into canonical form as described in section \ref{sec:specialcases}. After this, one can apply the ideas of \cite{Henn:2014qga,Lee:2014ioa} to the last row and obtain a canonical form for the full differential equation.

Instead, we choose to make use of the ability of our algorithm to test for further suitable candidates. 
Heuristic rules for finding UT integrals from ref.~\cite{Henn:2013fah} suggest that doubling a propagator is promising for this type of integral with off-shell triangle subintegrals. Looking at figure \ref{fig:P4L4P}, we see that there are two inequivalent ways of doubling one propagator.
 Checking both with our algorithm, we find that each of them is a UT integral (up to an overall normalization in $\eps$) but that only one of them has 19 linearly independent derivatives:
\begin{equation}
 f_1=g_1=\eps^7(1 + x)G_{1,1,1,1,1,1,1,2,0,0,0,0,0,0,0,0,0,0,0,0,0,0}
\end{equation}
Note that the factor of $(1+x)$ can easily be found through testing possible factors or by integrating out the $\eps^0$ part of $\tilde{b}_0$, as described in section \ref{sec:equsolve}.
Using this integral as our starting point, our algorithm takes less than a minute to find the canonical basis.


Solving the DE \eqref{decanonicalsinglevar} in terms of iterated integrals as in \eqref{equ:iteratedIntegralSol} is then straightforward. The alphabet, $\vec{\alpha}=\{x,1+x\}$, suggests that the solution can be written in terms of harmonic polylogarithms (HPLs) \cite{Remiddi:1999ew} with indices $\{0,-1\}$ only.

The boundary vector $\vec{g}(x_0,\eps)$ is determined, up to the overall normalization, by using the existence of a UV finite basis, see \cite{Chicherin:2018mue}. The normalization can be fixed by explicitly calculating an integral of our choice, which we took to be $G_{1,0,1,0,1,1,1,1,0,0,0,0,0,0,0,0,0,0,0,0,0,0}$. The result is given in an ancillary file. E.g.\ at $s=-1, t= -x$, the first few orders of the scalar integral are
\begin{align}
&  \epsilon^{6} (1-5 \epsilon)(1-6 \epsilon) \, G_{1,1,1,1,1,1,1,1,0,0,0,0,0,0,0,0,0,0,0,0,0,0}= 
  5 \epsilon^5 \zeta_5 + 
\epsilon^6  \left\{ 
 \frac{\pi^6}{30} + 3 \zeta_3^2 \right.   \nn \\
 & \hskip.3cm  + \frac16 \pi^4  H_{-1, -1}( x) + \frac15  \pi^4 \,  H_{-1, 0}(x) - 
 \frac{8}{45} \pi^4 \,  H_{0, -1} ( x)  \nn \\
&  \hskip.3cm  + \frac{2}{3} \pi^2 \,  H_{-1, -1, 0, 0} (x) 
 -  2 \pi^2 \,  H_{-1, 0, 0, -1} (x) + \frac23 \pi^2\,  H_{-1, 0, 0, 0} (x)  \nn \\
 & \hskip.3cm   -  \frac23 \pi^2\,  H_{0, -1, -1, 0} (x) + 2 \pi^2 \,  H_{0, 0, -1, -1} ( x) - 
 \frac23 \pi^2 \,  H_{0, 0, -1, 0} (x)  \nn \\
 & \hskip.3cm  + 4 \, H_{-1, -1, 0, 0, 0, 0}  (x) - 
 4 \, H_{-1, 0, 0, -1, 0, 0}( x)  \nn\\ 
&  \hskip.3cm  - 4\,  H_{0, -1, -1, 0, 0, 0} ( x) + 
 4 \, H_{0, 0, -1, -1, 0, 0} ( x) \nn \\
& \hskip.3cm  \left.  - \frac43  \pi^2 \zeta_3   \,  H_{-1} ( x)  - 
 4 \zeta_3 \, H_{0, -1, 0} ( x)  + 4 \zeta_3 \, H_{0, 0, -1}  (x) 
   -  20\zeta_5 \,  H_{-1}  (x )   \right\} \,. 
\end{align}
We have checked explicitly at $x=0$, that the result for the scalar integral agrees with HyperInt \cite{Panzer:2014caa} up to transcendental weight 7. 
The result has also been checked  numerically with pySecDec \cite{Borowka:2017idc} for $x=0, x=0.5$ and $x=2$.


\subsection{Canonical form for non-planar four-loop sector with 17 master integrals}
\label{sec:cusp}

\begin{figure}[t]
\centering
	\begin{center}
	\includegraphics[width=0.3\columnwidth]{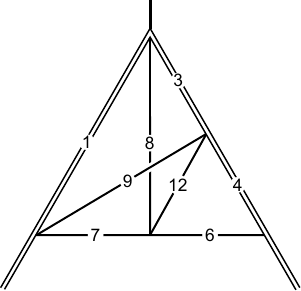}
		\end{center}
\caption{Non-planar four-loop integral appearing in the calculation of the cusp anomalous dimension.
The sector shown has $17$ coupled master integrals, according to \textsc{FIRE6}.
\label{fig:Topo6}
}
\end{figure}

Here we discuss one of the most complicated applications of our algorithm.
In the previous cases, the maximal size of individual sectors (i.e., the number of coupled integrals) was three in section \ref{sec:threeloop},
and twelve in section \ref{sec:fourloop}. In contrast, here we will apply the algorithm to a case of a sector with $17$ coupled master integrals.
It is shown in figure \ref{fig:Topo6}. 
The definition of the integral family is
\begin{equation}
 \begin{aligned}
 D_1&=1-2k_1\cdot v_1, &D_2&=1-2k_2\cdot v_1, &D_3&=1-2k_2\cdot v_2,\\
 D_4&=1-2k_3\cdot v_2,\qquad &D_5&=-k_1^2,\qquad &D_6&=-k_3^2,\\
 D_7&=-k_4^2,\qquad &D_8&=-(k_1-k_2)^2,\qquad &D_9&=-(k_1-k_4)^2,\\
 D_{10}&=-(k_2-k_3)^2,\qquad &D_{11}&=-(k_3-k_4)^2,\qquad &D_{12}&=-(k_1-k_2+k_3-k_4)^2,\\
 D_{13}&=-(k_2-k_4)^2,\qquad &D_{14}&=-(k_2-k_3-k_4)^2,\qquad &D_{15}&=1-2k_4\cdot v_1,\\
 D_{16}&=1-2k_4\cdot v_2, &D_{17}&=1-2k_3\cdot v_1, &D_{18}&=1-2k_1\cdot v_2,
 \end{aligned}
\end{equation}
where we consider the sector
$G_{1,0,1,1,0,1,1,1,1,0,0,1,0,0,0,0,0,0}$. 
Inspecting the differential equation for the cut integral, we identify the alphabet of the sector to be $\{x,1+x,1-x\}$, where $2v_1\cdot v_2=x+1/x$.

The algorithm needs one UT integral to start with. We used the following procedure to find this integral:
\begin{enumerate}
 \item We use heuristic rules to find likely UT candidates (see \cite{Henn:2013fah,Grozin:2015kna} for more information).
 \item We use our algorithm to test the UT property for each integral individually. If only the appropriate normalization factor is missing, we find it in the same ways as mentioned in the previous subsection.
\end{enumerate}
Following this procedure, we found the following integral to be UT on the maximal cut:
\begin{align}
\label{eq:3.10}
\eps^6\left(\frac{1 - x^2}{x}\right)^2 G_{1, 0, 1, 1, 0, 1, 1, 2, 2, 0, 0, 1, 0, 0, 0, 0, 0 ,
     0} \,. 
\end{align}
Starting from \eqref{eq:3.10}, our algorithm finds a UT basis in less than two minutes.

\subsection{Four-variable example: Non-planar double pentagon integrals}
\label{sec:fivevariable}

\begin{figure}[t]
    \begin{center}
	\includegraphics[width=0.3\columnwidth]{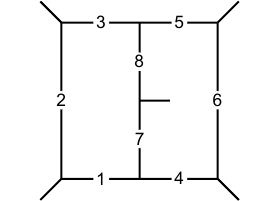}
		\end{center}
\caption{The non-planar double-pentagon integral family.}
\label{fig:Pentagon}
\end{figure}

Here we apply the algorithm to 
the cutting-edge example of a coupled system of two-loop five-point integrals, whose differential equation depends on four dimensionless variables
\cite{Chicherin:2018old,Abreu:2018aqd}.
Consider the non-planar double-pentagon integral family figure \ref{fig:Pentagon}, where the inverse propagators are,
\begin{equation}
 \begin{aligned}
 D_1&= -k_1^2 ,  &D_2&=- (k_1-p_1)^2 , & D_3&= - (k_1-p_1-p_2)^2,\\
 D_4&= -k_2^2 ,\quad &D_5&= -(k_2+p_4+p_5 )^2 ,\quad &D_6&= -(k_2+p_5)^2,\\
 D_7&= -(k_1- k_2)^2,\quad &D_8&= -(k_3+ k_1-k_2)^2,\quad &D_9&= - (k_1+p_5)^2,\\
 D_{10}&=-(k_2-p_1)^2,\quad &D_{11}& =(k_2-p_1-p_2)^2,
 \end{aligned}
\end{equation}
We focus on the top sector $G_{1,1,1,1,1,1,1,1,0,0,0}$ which contains 9 master integrals. 
We start with an ansatz for the alphabet containing 31 letters $\{ W_i \}$, suggested by \cite{Gehrmann:2018yef,Chicherin:2017dob}.
The five-point external kinematics can be parametrized via (a variation of) momentum-twistor variables ${\bom b} = \{ b_1, \ldots, b_5 \}$, which rationalize all letters of the alphabet, 
see e.g.\ \cite{Badger:2013gxa}.
\begin{align}
s_{12} &= b_1, &s_{23} &= b_1 b_4, &s_{34} &= \frac{b_1 (1 + b_3) b_4}{b_2} - b_1 b_3 (1-b_5),\\
s_{45} &= b_1 b_5, &s_{15} &= b_1 b_3 (b_2 - b_4 + b_5) & &
\end{align}
where $b_1$ sets the overall kinematic scale.  
 The differential equation depends on four dimensionless variables $b_2 ,\dots , b_5$. 
The algorithm takes derivatives with respect to one preferred variable. 
It is convenient for us to choose this variable to be  $b_2$. 
 Given the ansatz $\{ W_i\}$, the canonical partial differential matrix $B_2 ({\bom b}, \epsilon) \equiv \frac{\partial }{ \partial b_2 }B $ depends on 22 independent letters,
 \begin{align}
 B_2 ({\bom b}, \epsilon) \equiv  \epsilon \sum_{l= 1}^{22} \, \mathbf{m}_{l} \,  \frac{\partial}{\partial b_2 }  \hskip-.05cm  \log {\alpha_l ({\bom b } ) }  \,. 
 \end{align} 
To proceed, we need to select a suitable UT integral as an input to the algorithm. 
In order to investigate the parity dependence of the differential equations, 
we tested the algorithm starting with both a parity-even and parity-odd integral taken from the canonical basis given in \cite{Chicherin:2018old}. 
 Fixing the value of $b_3, b_4, b_5$ to be certain constants, we execute the algorithm to find a solution for $\mathbf{m}_1, \ldots, \mathbf{m}_{22}$, 
 and hence for $ B_2 ({\bom b}, \epsilon)$.   
 In order to determine the transformation matrix $T ({\bom b},\eps)$, as defined in \eqref{transformT}, normally one would compute the $\Psi$ and $\Phi$ matrix with full analytic depedence on all variables. 
This is the most time-consuming step of the algorithm since 
 taking higher derivatives generates rational coefficients that depend on high-degree polynomials. 
 In this case, we find it more convenient to first compute the matrices $B_i  \equiv \frac{\partial}{ \partial b_i} B, i = 1, \ldots ,5$,  which  allow us to determine the whole differential in \eqref{conclusion-multi-variable} analytically.  
 
We proceed in the following way: 
\begin{itemize}
\item Set $b_4$ and $b_5$ to constants. 
Compute $\Psi_2, \Phi_2$ and hence the $T$ matrix with analytic dependence on $b_3$. 
Transform the original partial differential equation in $b_3$ into canonical form. 
Thus we obtain $B_3$ where all other variables are set to constants.
\item  
 By matching onto the ansatz for the relevant  alphabet letters,  we extract the corresponding matrix-residues in $B_3$.  
 In this way we reconstruct $B_3$ with analytic dependence on all variables.
\item Repeat the above procedures to find $B_4 ({\bom b} , \epsilon), B_5 ({\bom b} , \epsilon)$. 
\item Construct an intermediate integral basis $\vec{h}$ to relate the canonical basis $\vec{g}$ to the  initial basis $\vec{f}$.  
For example, we set $\vec{h} = (f_1, \frac{\partial}{ \partial b_2 } f_1, \frac{\partial^2}{ \partial b_2^2 } f_1,
\frac{\partial}{ \partial b_3 } f_1, \frac{\partial^2}{ \partial b_3^2} f_1 , \frac{\partial}{ \partial b_4 } f_1,\frac{\partial^2}{ \partial b_4^2 } f_1,
\frac{\partial}{ \partial b_5 } f_1, \frac{\partial^2}{ \partial b_5^2 } f_1) $ which contains only  first- and second-order partial derivatives. 
\item Find the linear transformation between $\vec{h}$ and $\vec{g}$ through the  $B_i$ matrices. 
 Likewise, work out the linear relations between  $\vec{h}$ and the initial basis $\vec{f}$ through the original differential equations. 
 The way we construct $\vec{h}$ guarantees that these relations can be easily obtained analytically.  
\item Compute the  transformation matrix $T({\bom b}, \epsilon) $ between  $\vec{g}$ and $\vec{f}$ through their relations to $\vec{h}$. 
\end{itemize}

Starting from either a parity-even or parity-odd UT integral,  the algorithm finds the solution 
to transform the differential equations on the maximal cut into canonical form, which depend on 17 letters $W_i , i \in \{ 1 ,\ldots, 5,11,16 ,\ldots, 20,26 ,\ldots, 31 \} $.
 The total running time for solving the double-pentagon example is of the order of minutes.  
 As a starting point, the algorithm needs to know one UT integral. 
 This information can be obtained from D-dimensional Baikov representation \cite{Chicherin:2018old}.
  Alternatively,  one could start with a \emph{dlog} integral in six dimensions \cite{Abreu:2018aqd}. 
  Compared with the methods in the literature, our algorithm requiries minimum input from the integrand analysis \cite{WasserMSc}. 
This feature makes the algorithm particularly suitable for dealing with multi-variable problems. 
    
The performance of our algorithm on all examples of this section is summarized in table \ref{tab:benchmarks}.

\begin{table}[t!]
\centering
\begin{tabular}{ |c||c|c|c||c|c| } 
\hline
 type of problem & \#MI & \#vars & \#letters & time & mem. \\
  &  &  &  & [min] & [MB] \\
 \hline
 \hline
three-loop four-point ladder &  26 | 3 &  1 &  2 & 2  &  330  \\
\hline
three-loop four-point tennis court &  41 | 3 &  1 &  2 & 34  &  1710  \\
\hline
four-loop four-point crossed box &  19 | 12 &  1 &  2 &  1 &  240  \\
\hline
non-planar four-loop HQET &  17 | 17 &  1 &  3 & 2  &  390  \\
\hline
non-planar two-loop five-point &  9 | 9 &  4 &  17 &  5  &  510 \\
\hline
\end{tabular}
\caption{Approximate evaluation time and memory usage of the different examples on a desktop computer with twelve logical CPU cores. The second column shows the total number of master integrals, as well as the maximum sector size. The third column shows the number of dimensionless variables and the fourth column gives the number of (relevant) letters in the alphabet. }
\label{tab:benchmarks}
\end{table}

\section{Public implementation}
\label{sec:MathImplement}

We provide a \textsc{Mathematica} package  INITIAL (an INitial InTegral ALgorithm)  which utilizes \textsc{FiniteFlow} \cite{Peraro:2019svx} to perform operations of our algorithm over finite fields.  
%
%
The package is  publicly available at
\begin{equation*}
\text{\href{https://github.com/UT-team/INITIAL}{\texttt{https://github.com/UT-team/INITIAL}}}
\end{equation*}
It relies on the \textsc{FiniteFlow} library \cite{Peraro:2019svx} and its dependencies. The examples mentioned in the previous section can also be downloaded from the same repository.

\section{Conclusion and outlook}
\label{sec:conclusionandoutlook}

The automated calculation of Feynman integrals in quantum field theory is of considerable interest
to the scientific community. In this paper, we developed further an idea due to \cite{Hoschele:2014qsa}, which in turn relies
on the method of canonical differential equations \cite{Henn:2013pwa}. 

In other approaches, one needs a full set of uniform weight integrals to obtain the canonical differential equations.
In the new approach, only one such integral is needed: The remaining ones are obtained from the former algorithmically,
if such a basis exists. 
As a corollary, the new approach provides an algorithm to test whether the candidate integral has uniform weight.
 A necessary condition is given in \eqref{UTcondition}.  
 Moreover, the existence of a canonical transformation verifies the uniform weight
property of the candidate integral.

We expressed the necessary equations in matrix form. The equations are easily handled, and we explained how to
solve them systematically. We found that this implementation can be readily used for state-of-the-art problems.
We used it to find the canonical form of complicated systems of differential equations, e.g. with 17 coupled master integrals
in one integral sector (on the cut).  We explained how the algorithm deals with multi-variable differential equations in an efficient way. 
For demonstration, we applied it to a cutting-edge two-loop five-point example.  Given one UT integral, the algorithm finds the canonical transformation for the non-planar double-pentagon  integrals on the maximal cut, which depends on four dimensionless variables.




Our work opens up several interesting directions for further developments:

{\underline{Canonical forms for elliptic polylogarithms:}}
The idea of the canonical form of differential equations has also been explicitly applied for elliptic polylogarithms. 
Specifically, there are two approaches that seem promising in this regard. Firstly, it has been argued \cite{Henn:2014qga} (see also \cite{Mizera:2019vvs}) that a pre-canonical form should exist for all Feynman integrals, of the following type,
\begin{align}
\diff {\vec{g}}({{x}}) = \left[ \diff A_{0}({{x}}) + \epsilon \diff A_{1}({{x}}) \right] {\vec{g}}({{x}})\,,
\end{align}
where the matrices $A_{0}$ and $A_{1}$ only involve logarithms (i.e., the fuchsian property of the
differential equations is manifest for all singular points).
In the polylogarithmic case, `integrating out' the $A_{0}$ part can be done using algebraic functions only,
but in the elliptic case (and beyond) this leads to special functions.
In the latter case, it has been shown explicitly that allowing non-algebraic transformation matrices 
one finds the canonical form of \cite{Henn:2013pwa} also in the elliptic case,
\begin{align}
\diff {\vec{f}}({{x}}) =   \epsilon \diff A({x})  {\vec{f}}({{x}})\,,
\end{align}
where $A({x})$ contains functions beyond logarithms \cite{Adams:2018yfj,Broedel:2018qkq}.

{\underline{Application to finite integrals:}} 
In most applications, for example when computing finite cross sections, or scaling dimensions of operators, one is ultimately 
interested in a finite, four-dimensional quantity. In some situations, it is possible to explicitly separate divergent parts of the calculation
and express the remainder in terms of manifestly finite integrals.
 It is therefore interesting to apply the method for finite integrals \cite{Caron-Huot:2014lda}. 
The latter occur e.g. when making the infrared properties of
scattering amplitudes manifest. 
Indeed, it is very natural within the {\it dlog} integrand approach to 
classify integrals according to their infrared properties, with the infrared finite integrals typically
being the most interesting ones \cite{ArkaniHamed:2010gh,Henn:2013pwa}. 
Moreover, finite integrals occur  frequently when dealing with problems
with several mass scales. 

The main simplification when dealing with finite integrals is that the weight expansion truncates: there is a
maximal weight occurring in the calculation.
The way this happens in practice is that the differential equation matrix becomes nilpotent.
Moreover, the number of master integrals reduces in this case \cite{Remiddi:2013joa} (some integrals decouple),
and it has also been observed that the required function alphabet may simplify.
Due to these simplifications, in ref.\ \cite{Caron-Huot:2014lda} a full massive three-loop calculation was possible.
We find it promising to combine this method with our new approach.

{\underline{Uniform weight as guiding principle for recurrence relations:}} 
Finally, let us mention that  
Feynman integrals satisfy dimensional recurrence relations \cite{Tarasov:1996br}.
For single-scale integrals, where differential equations can only be applied indirectly \cite{Henn:2013nsa},
they are one promising method to evaluate Feynman integrals. See e.g.\ \cite{Lee:2019lsr,Magerya:2019csf}
for recent applications. We find it likely that the uniform weight information of a single UT integral provides important input for that method as well,
and provides a useful organizing principle for the calculation.

\medskip
We wish to close with a discussion on the role played by $\mathcal{N}=4$ super Yang-Mills in our understanding of Feynman integrals. 
In many calculations of physical quantities in that theory it was observed that the answer typically is given by UT functions. Whether or
not this is true in general is an open question.
One may wonder whether the integrals encountered in $\mathcal{N}=4$ super Yang-Mills are simpler or more complicated, with respect to QCD.
Of course, when talking about full amplitudes or finite quantities in $\mathcal{N}=4$ super Yang-Mills, the latter often have additional symmetries or hidden properties, and hence are definitely special. However, what about individual Feynman integrals in dimensional regularization? 
The answer is that generic QCD integrals are just as nice, or simple, as the ones in $\mathcal{N}=4$ super Yang-Mills,
at least if one organizes them in a suitable way, as we have learned in the course of the last ten years.
With the present work, it becomes clear that a stronger statement is possible: 
Not only are generic integrals as simple as the ones in $\mathcal{N}=4$ super Yang-Mills, 
but in fact they can all be obtained from the knowledge of the former!

\acknowledgments

We thank Tiziano Peraro for support with \textsc{FiniteFlow} and Simone Zoia for testing our implementation. 
JMH thanks the organizers of the workshop {\it MathemAmplitudes 2019: Intersection Theory and Feynman Integrals} (Padova, December 18-20,2019), where this work was presented, for their invitation.
This research received funding from the European Research Council (ERC) under the European Union’s Horizon 2020 research and innovation programme,
{\it Novel structures in scattering amplitudes} (grant agreement No 725110).




 \bibliographystyle{JHEP} 
 \bibliography{BibFile} 

\end{document}